\documentclass[12pt]{article}
\RequirePackage[OT1]{fontenc}
\RequirePackage{amsthm,amsmath}
\usepackage{breqn}
\usepackage{soul}
\usepackage[numbers,sort&compress]{natbib}
\RequirePackage[citecolor=blue,urlcolor=blue]{hyperref}
\usepackage{hyperref}
\usepackage{amsfonts}
\usepackage{amssymb}
\usepackage{graphicx}
\usepackage{setspace}

\numberwithin{equation}{section}
\newcommand{\bxi}{{\boldsymbol \xi}}

\newcommand{\bet}{{\boldsymbol \eta}}

\newcommand{\bOm}{{\boldsymbol \Omega}}

\newcommand{\bLambda}{{\boldsymbol \Lambda}}

\def\b1{{\mathbf 1}}
\def\b0{{\mathbf 0}}
\def\bw{{\boldsymbol w}}
\def\bA{{\boldsymbol A}}
\def\bB{{\boldsymbol B}}

\def\bk{{\boldsymbol k}}
\def\bK{{\boldsymbol K}}
\def\bv{{\boldsymbol v}}

\def\bZ{{\boldsymbol Z}}
\def\bp{{\boldsymbol p}}
\def\bP{{\boldsymbol P}}
\def\bE{{\boldsymbol E}}
\def\bQ{{\boldsymbol Q}}

\def\bx{{\boldsymbol x}}

\def\bH{{\boldsymbol H}}
\def\bM{{\boldsymbol M}}

\def\bs{{\boldsymbol s}}
\def\bn{{\boldsymbol n}}

\def\bU{{\boldsymbol U}}
\def\bZ{{\boldsymbol Z}}

\def\cE{{\mathcal E}}

\def\cH{{\mathcal H}}

\def\bbH{{\mathbb H}}

\def\bbI{{\mathbb I}}

\graphicspath{%
    {converted_graphics/}
    {/}
}

\begin{document}

\title{Canonical equivalence of a charge in a time\\dependent, spatially-homogeneous electromagnetic\\ field to a time-dependent perturbed oscillator}
\author{Henryk Gzyl\\
Centro de Finanzas IESA, Caracas, Venezuela.\\
 henryk.gzyl@iesa.edu.ve}
\date{}
 \maketitle

\setlength{\textwidth}{4in}

\vskip 1 truecm
\baselineskip=1.5 \baselineskip \setlength{\textwidth}{6in}
\begin{abstract}
Here we prove that the classical (respectively, quantum) system, consisting of a particle moving in a static electromagnetic field, is canonically (respectively, unitarily) equivalent to a harmonic oscillator perturbed by a spatially homogeneous force field. This system is canonically and unitarily equivalent to a standard oscillator. Therefore, by composing the two transformations we can integrate the initial problem. Actually, the eigenstates of the initial problem turn out to be entangled states of the harmonic oscillator.

When the magnetic field is spatially homogeneous but time-dependent, the equivalent harmonic oscillator has a time-varying frequency. This system can be exactly integrated 
only for some particular cases of the time dependence of the magnetic field.

The unitary transformations between the quantum systems are a representation of the canonical transformations by unitary transformations of the corresponding Hilbert spaces.
\end{abstract}


\section{Introduction and Preliminaries} 
We first establish the canonical equivalence between the Hamiltonians describing the motion of a particle in a static electromagnetic field to that of a classical harmonic oscillator. Then we show how to implement the canonical transformations by unitary transformations between the corresponding Hilbert spaces. This is done in two steps, first, we map the particle moving under the action of the electromagnetic fields into a harmonic oscillator subject to a spatially constant, but time-dependent force, and then we show that this system is equivalent to a simple harmonic oscillator.

To do this, we put together two separate lines of work, in which each of the two steps was carried out separately. See \cite{Gz1} and\cite{Gz2}. The second step was considered only in one dimension. But it contains two subcases that are useful here, as well as many references to previous work that are not mentioned here.

Let us establish some notational conventions. Vectors will be either 3-dimensional or 6-dimensional column vectors written in boldface. To refer to the components on $\bv,$ we write $\bv=(v_1,v_2,v_3)^\dag,$ where the superscript ``$\dag$'' will always mean transpose of the corresponding object. We also need $\bv=(\hat{\bv},v_3)^\dag$ where $\hat{\bv}=(v_1,v_2)^\dag$ will stand for the 2-vector consisting of the first two components of $\bv.$ By $\langle\bv,\bw\rangle$ (resp. $\bv\times\bw$) we denote the usual scalar (resp. vector) product of the two vectors.

The classical dynamics of the three systems are described by:

\begin{gather}
H_1(\bx,\bp) = \frac{1}{2m}\big(\bp-\bA(\bx)\big)^2 -q\langle\bx\bE\rangle.\label{ch1}\\
H_2(\bQ,\bP) = \frac{1}{2m}\|\bar{\bP}\|^2 +\frac{1}{2m}P_3^2  + \frac{1}{2}\omega^2\|\bar{\bQ}\|^2 - q\langle\bQ,\bE(t)\rangle.\label{ch2}\\
H_3(\bxi,\bet) = \frac{1}{2m}\|\bar{\bet}\|^2 + \frac{1}{2m}\eta_3^2 + \frac{1}{2}\omega^2\|\bar{\bxi}\|^2 .\label{ch3}
\end{gather}

The time evolution of the corresponding quantum systems is determined by the operators
\begin{gather}
\bH_1 = \frac{1}{2m}\big(-i\hbar\nabla-\bA(\bx)\big)^2 -q\langle\bx\bE\rangle.\nonumber\\
\bH_1 = -\frac{\hbar^2}{2m}\Delta_{\bx} -\frac{i\hbar}{2}\langle\bOm\bar{\bx},\nabla_x\rangle +  \frac{m\omega^2}{2}\langle\bar{\bx},\bar{\bx}\rangle -q\langle\bx\bE\rangle \label{qh1}\\
\bH_2 = -\frac{\hbar^2}{2m}\Delta_x  +\frac{1}{2}\omega^2\|\bar{\bQ}\|^2 - q\langle\bQ,\bE(t)\rangle.\label{qh2}\\
\bH_3  = -\frac{\hbar^2}{2m}\Delta_{\bxi}  +\frac{1}{2}\omega^2\|\bar{\bxi}\|^2.\label{qh3}
\end{gather}
The explanations of the notations come up a few lines below.
In each set, the first Hamiltonian describes the motion of a particle of charge $q$ and mass $m$ in a static electromagnetic field. The second describes the motion of a particle under the action of a linear restoring force plus a spatially constant but time-dependent force, whereas the third describes a particle under the action of a planar restoring force. Note as well that if $\bE$ were absent, then $H_2$ and $H_3$ coincide. Ditto for their quantized versions. To obtain the version of $\bH_1$ in the second line from the first, note that when acting on functions of $\bx, $ we have:
$$\langle\bOm\bx,\nabla_{\bx}\rangle + \langle \nabla_{\bx},\bOm\bx,\rangle = 2\langle\bOm\bx,\nabla_{\bx}\rangle + div(\bOm\bx) = 2\langle\bOm\bx,\nabla_{\bx}\rangle$$
because $div(\bOm\bx)=tr(\bOm)=0.$
Above we used the symbol $\bOm$ to denote the cross product matrix for $\bB,$
 that is $\bOm\bx=\bB\times\bx$ for any $\bx.$ Except for the presence of the electric field, the classical (and quantum) equivalence of \eqref{ch1} and \eqref{ch2} (respec. \eqref{qh1} and \eqref{qh2}) is, essentially the subject matter of \cite{Gz1}. Here we extend the result to the current setup. The thrust of \cite{Gz2} is to provide a one-dimensional equivalence of \eqref{ch2} to \eqref{ch3} (respec. \eqref{qh2} to \eqref{qh3}), and to examine the possible global phase that appears when the particle under a restoring force is perturbed by a time-dependent, but spatially homogeneous force. The extension considered here implies that the presence of a magnetic field does not induce a global change of phase unless there is also an electric field present. 
 
  We also consider the case in which the magnetic field is spatially homogeneous, but time-dependent. We consider two cases: First, the magnetic field is time-dependent, but its direction is fixed, and second, the magnetic field rotates about a fixed axis. In the first case, the system is equivalent to a harmonic oscillator subject to an external, spatially constant, but time-dependent force.  As we shall see below, the two cases are equivalent to a harmonic oscillator with a time-varying frequency. In the first case, the frequency is time-dependent, but is the same for all components of the oscillator. See \cite{MW} for example. The equation of motion is of the Hill type. For a discussion in the applied mathematics literature, not overlapping much of the literature in physics see \cite{MW} for example. In the physical literature  see \cite{K}, \cite{LR}, \cite{GC}, \cite{EG}, \cite{L}. For a group theoretical study of quadratic Hamiltonians with time-dependent coefficients, see \cite{W}. Both cite{SK} and \cite{P} use canonical transformations in a way totally unrelated to ours. For a review of the Ermakov invariant, which is an approach used in several of the works just cited, see \cite{L}.
 This invariant is an ingenious way to deal with the oscillator with a time-dependent frequency, but it works for particular time dependencies.  For other approaches to the problem discussed here see \cite{dJ}, \cite{EB}, \cite{DPR}. The description of the motion of a single particle in a static electromagnetic field is important for the computation of the magnetic moment of the electron. See \cite{BG} for example.
  
We devote the remainder of this section to introducing more notations and establishing some preliminary results. Then in Section 2, we establish the classical equivalences between the Hamiltonians, We add that the equivalence has been noted before, and it is already a textbook matter, but the contribution here is to present the equivalence as a canonical transformation.  In Section 3, establish the equivalences of the quantized version of the systems, by making use of the generating functions to define the unitary that realizes the equivalence between these systems.  Once the equivalence of the original system to a harmonic oscillator in an external field has been established, we relate the energy eigenstates of the original system to those of the harmonic oscillator. The result is that the eigenstates of a charged particles in a static electromagnetic field are entangled states of the simple harmonic oscillator. 

 In Section 5 we consider two variations on the theme of a particle moving in a spatially constant but time-dependent electromagnetic field.  There we establish that the techniques in Section 2 lead to a harmonic oscillator with time-depending frequency. This system has been studied considerably.  As mentioned above, in the general case, it can only be dealt with using approximations.

\subsection{Notations and preliminary results}
When $\bB$ is constant, $\bA(\bx)=\bB\times\bx/2 =\bOm\times\bx.$  When $\bB=B\hat{\bk},$  then:
\begin{equation}\label{ang1}
\bOm = \begin{pmatrix}0 & -\omega &0 \\
                \omega & 0 & 0\\
								0 & 0 & 0
				\end{pmatrix} = \begin{pmatrix} \bOm_0 & \b0\\	
			\b0^t & 0\end{pmatrix}\;\;\;\mbox{with}\;\;\;\bOm_0=\begin{pmatrix} 0 & -\omega\\
				                                 \omega & 0\end{pmatrix}
\end{equation}	
			
where we put $\omega=\frac{qB}{mc}$ for the standard cyclotron frequency. The vector $\b0$ is two dimensional zero vector, and the superscript ``t'' stands for the transpose of the indicated object. Also, keep in mind that $\bOm_0^t\bOm_0=\omega^2\bbI.$ Below we make extensive use of the fact that
\begin{equation}\label{rot1}
R(t) = e^{t\bOm} = \begin{pmatrix}e^{t \bOm_0} & \b0\\	
			\b0^t & 1\end{pmatrix}\;\;\;\mbox{with}\;\;\;
			e^{t\bOm_0}=\begin{pmatrix} \cos(\omega t) & \sin(\omega t)\\
				                                \sin(\omega t) & \cos(\omega t)\end{pmatrix}
\end{equation}
stands for a rotation matrix about the $z-$axis with constant angular speed $\omega.$

To establish the canonical equivalence between \eqref{ch2} and \eqref{ch3}, we need to compute the trajectories of the motion described by the Hamiltonian \eqref{ch2}. Notice that the three degrees of freedom are separated. We have
$$\frac{1}{2m}P_i^2 +  \frac{1}{2}\omega^2 Q_i^2  - q Q_iE_i(t),\;\;\;i=1,2$$
and
$$\frac{1}{2m}P_3^2 - q Q_3E_3(t).$$
The last one is obtained from the former setting $\omega=0.$ Temporarily dropping the reference to the label of the coordinates, the Hamilton equations of motion are:

\begin{equation}\label{H1}
\begin{aligned}
&\frac{d}{dt}{Q \atopwithdelims ( ) P} = {\frac{\partial H}{\partial P} \atopwithdelims ( ) -\frac{\partial H}{\partial Q}} = {P \atopwithdelims ( ) -\omega^2 Q+k(t)} =
\left(\begin{array}{cc}
	0 & 1\\
-\omega^2 & 0\end{array}\right){Q \atopwithdelims ( ) P} +{0\atopwithdelims ( )k(t)}\\
& =\bbH_0{Q \atopwithdelims ( ) P} +{0\atopwithdelims ( )k(t)}.	
\end{aligned}
\end{equation}
The initial conditions are $Q(0)=Q_0, P(0)=P_0.$ We put $k(t)=eE(t)$ for short. The equations of motion of the unperturbed oscillator are obtained by setting $k(t)=0,$ and the solution for $(Q_3, P_3)$ is obtained by setting $\omega=0.$ Write $Z(t)=(Q(t),P(t))^\dag.$ The solution of the system \eqref{H1} is:
\begin{equation}\label{sH2}
Z(t) = U(t)Z(0) + \int_0^tU(t-s)\bk(s)ds = Z_h(t) + Z_{nh}(t).
\end{equation}
where the matrix $U(t)$ is given by:
\begin{equation}\label{sH2.1}
U(t) = \left(\begin{array}{cc}
	cos(\omega t) & \frac{1}{\omega}sin(\omega t)\\
-\omega sin(\omega t) & cos(\omega t)\end{array}\right),
\end{equation}
Clearly $U(t)$ satisfies $U(t+s)=U(t)U(s)$ or $U(t-s)=U(t)U(-s)$ for all $s,t.$ In the last term of \eqref{sH2}, $\zeta_h(t)$ denotes the first term in the middle and subscript $h$ stands for {\it homogeneous} and, $nh$ stands for {\it non-homogeneous} in the last term. 

Note that $Z_{nh}(t)$ is just the particular solution to (\ref{H1}) with zero initial conditions, and it describes the motion of the origin of the coordinate system. therefore, we might think of \eqref{sH2} as the position of the particle with respect to a system whose origin of coordinates moves according to $Z_{nh}.$ Also, $Z_h(t)=Z(t)-Z_{nh}(t)$ describes the motion of a simple harmonic oscillator, which is consistent with the fact that

$$\langle\big(Z(t)-Z_{nh}(t)\big),\bbH_0\big(Z(t)-Z_{nh}(t)\big)\rangle=\mbox{constant}=\langle Z(0),\bbH_0Z(0)\rangle.$$

This follows readily from the fact that 
\begin{equation}\label{inv1}
U^\dag(t)\bbH_0 U(t) = \bbH_0.
\end{equation}

To go from this to the 3-dimensional case, let us introduce the following more compact notations
\begin{gather}
\bar{Z}_1={Q_1\atopwithdelims ( ) P_1},\;\;\bar{Z}_2={Q_2\atopwithdelims ( ) P_2},\;\;\bar{Z}_3={Q_3\atopwithdelims ( ) P_3},\label{comp1}\\
\bZ = ( \bar{Z}_1, \bar{Z}_2, \bar{Z}_3)^\dag. \label{comp2}
\end{gather}
As at the beginning, $\bQ=(\bar{\bQ}^\dag,Q_3)^\dag=(Q_1,Q_2,Q_3)^\dag$ and $\bP=(\bar{\bP}^\dag,P_3)^\dag=(P_1,P_2,P_3)^\dag.$ To describe the solution to the full equation of motion, we introduce the following notations:

\begin{equation}
\bU(t)= \begin{pmatrix} U(t) &  0 & 0 & 0\\
					0 & U(t) & 0 & 0\\   
					0 &  0   & 1  & t\\
                                     0  &  0 &  0 & 1\end{pmatrix} 
\end{equation}

Here $U(t)$   is the $2\times2-$matrx introduced in \eqref{sH2.1}. With that, we have:

\begin{equation}\label{sH3}
\bZ = \bU(t)\bZ(0) + \int_0^t \bU(t-s)\bK(s)ds = \bZ_{h}(t)+\bZ_{nh}(t).
\end{equation}
Here we put $\bK(t) = \big((0,qE_1(t)), (0,qE_2(t)), ((0,qE_3(t))\big)^\dag.$ For the record, let us write explicitly what $Z_3(t)$ looks like. According to \eqref{sH2} we have:
\begin{equation}\label{sH4}
Z_3(t) = {Q_3(0)+P_3(0)t\atopwithdelims ( ) P_3(0)} + {\int_0^t\frac{\sin(\omega(t-s)}{\omega}qE_3(s)ds \atopwithdelims ( ) \int_0^t \cos(\omega(t-s) qE_3(s)ds} = Z_{3,h}(t) +Z_{3,nh}(t).
\end{equation}
It is up to the reader to verify that this reduces correctly to the solution when $\omega=0$ and/or $E_3(t)=$ constant .
                                   
\section{The classical equivalence of the Hamiltonians}
The the canonical transformation relating \eqref{ch1} to \eqref{ch2} is determined from the following generating function (see \cite{A} or \cite{G}):

\begin{equation}\label{CT1}
F(\bx,\bP,t) = \langle\bx,U(-t/2)\bP\rangle + A(t) = \langle\bar{\bx},e^{-t\bOm_2/2}\bar{\bP}\rangle +x_3P_3.
\end{equation}	

The transformation equations (change of variables) that \eqref{CT1} induces is:
\begin{gather}
\bar{\bQ} = e^{t\bOm_0/2}\bar{\bx};\;\;\bar{\bP}=e^{t\bOm_0/2}\bar{\bp}, \;Q_3=x_3, \;P_3=p_3, \;\bE(t)=R(t/2)\bE. \label{cteq1.1} \\
H_2(\bQ,\bP)  = H_1(\bx,\bp) + \frac{\partial F}{\partial t} = \frac{1}{2}\langle\bar{\bP},\bar{\bP}\rangle+ \frac{m\omega^2}{2}\langle\bar{\bQ},\bar{\bQ}\rangle + \frac{1}{2}P_3^2. \label{cteq1.2}
\end{gather}
		 
We used the fact that 	$\langle\bar{\bP},\bar{\bP}\rangle=\langle\bar{\bp},\bar{\bp}\rangle,$ $\langle\bar{\bQ},\bar{\bQ}\rangle=\langle\bar{\bx},\bar{\bx}\rangle,$ and that 
$$\partial F_2/\partial t =\frac{1}{2} \langle\bar{\bp},\bOm_0\bar{\bx}\rangle,$$
using \eqref{cteq1.2} after differentiating. In the new coordinates, we have a two-dimensional harmonic oscillator plus a free motion along the $Q_3-$axis.

To establish the equivalence of \eqref{ch2} to \eqref{ch3} we use the transformation generated by:
\begin{equation}\label{CT2}
F(\bQ,\bet,t) = \langle\big(\bQ-\bQ_{nh}(t)\big),\big(\bet + m\dot{\bQ}_{nh}(t)\big)\rangle+A(t)
\end{equation}
To begin with, this leads to the following change of variables:
\begin{equation}\label{cteq2.1}
\bxi = \bQ - \bQ_{nh},\;\;\;\bet = \bP - m\dot{\bQ}_{nh}.
\end{equation}
To obtain $H_3(\bxi,\bet)$, use $H_3=H_2+\partial F/\partial t,$ use \eqref{cteq2.1} and the fact that $m\ddot{\bQ}_{nh}=-m\omega^2\bQ_{nh}+q\bE(t),$ and require that $A(t)$  satisfies
\begin{equation}\label{req1}
\dot{A}(t) -\frac{m}{2}\langle\dot{\bQ}_{nh},\dot{\bQ}_{nh}\rangle +\frac{m\omega^2}{2}\langle\bQ_{nh},\bQ_{nh}\rangle - q\langle\bQ_{nh},\bE(t)\rangle = 0.
\end{equation}
This leads to \eqref{ch3}. What is perhaps interesting is that form  \eqref{req1} we obtain:
\begin{equation}\label{act}
A(t) = \int_0^t L(\bQ_{nh}(s),\dot{\bQ}_{nh}(s))ds,
\end{equation}
which happens to be the action along the curve $t \to \bQ_{nh}(t).$ Of course, $L$ is the Lagrangian function dual to $H_2.$

\section{Unitary representation of the canonical transformations}
As underlying state space for any of the three systems, we consider the space $\cH$ of square-integrable functions, and do not worry too much about matters related to the domain of the differential operators that come up. 

Given a state vector $|\psi\rangle,$ we denote its representation in coordinates or in momenta by $\psi(\bx)$ and $\psi(\bp).$ These two are related as usual, that is:
\begin{equation}\label{FT}
\psi(\bp) = \frac{1}{(2\pi)^{3/2}}\int e^{-i\langle\bp,\bx\rangle}\psi(\bx)d\bx,\;\;\;\psi(\bx) = \frac{1}{(2\pi)^{3/2}}\int e^{i\langle\bp,\bx\rangle}\psi(\bp)d\bp.
\end{equation}
Similarly for the other two pairs of conjugate variables: $(\bQ,\bP)$ and $(\bxi,\bet).$\\
Since all canonical transformations reduce to the identity at $t=0,$ we suppose that the state at $t=0$ is the same in all three cases regardless of the time evolution operator chosen. Since the simplest Hamiltonian is $H_3,$ we suppose that we know how to solve
\begin{equation}\label{SE1}
i\hbar\frac{\partial\psi(t) }{\partial t} = \bH_3\psi(t) \;\;\;\mbox{with}\;\;\; \psi(0) \;\mbox{given}.
\end{equation}
The idea is to define a representation of \eqref{CT2} by means of a time-dependent unitary transformation $U_t$, and prove that $\phi(t)=U_t\psi(t)$ satisfies
\begin{equation}\label{SE2}
i\hbar\frac{\partial\phi(t) }{\partial t} = \bH_2\phi(t) \;\;\;\mbox{with}\;\;\; \phi(0)=\psi(0).
\end{equation}
Similarly, we implement \eqref{CT1} by a unitary transformation $U_t$ -we use the same symbol and use the tags for the coordinates to tell each case apart, and prove that if $\phi(t)$ solves \eqref{SE2}, then $\varphi(t)=U_t\phi(t)$ solves
\begin{equation}\label{SE3}
i\hbar\frac{\partial\varphi(t) }{\partial t} = \bH_1\varphi(t) \;\;\;\mbox{with}\;\;\; \varphi(0)=\phi(0)=\psi(0).
\end{equation}
As the easiest equation to solve is \eqref{SE1}, we first show how to obtain the solution to \eqref{SE2} from the solution to \eqref{SE1}, and then how to obtain the solution to \eqref{SE3} from that of \eqref{SE2}.

So, let $\phi(t,\bQ)$ be a solution to \eqref{SE1} with initial condition $\psi_0(Q).$ The unitary version of \eqref{CT1} is defined by
\begin{equation}\label{QT1}
\psi(t,\bx) =  U_t\phi(t,\bx) = \frac{1}{(2\pi)^{3/2}} \int e^{iF(\bx,\bP,t)/\hbar}\phi(t,\bP)d\bP.
\end{equation}
Here $F(\bx,\bP,t)$ is given by \eqref{CT1}. Invoking \eqref{FT}, this reduces to the identity transformation at $t=0.$  A simple computation using \eqref{FT} yields
\begin{equation}\label{QT1.1}
\psi(t,\bx) =  \phi(t,R(t/2)\bar{\bx},x_3).
\end{equation}
The essential computation here is the following.
$$\frac{\partial }{\partial t}\psi(t,R(t/2)\bar{\bx},x_3) = \big(\frac{\partial \psi}{\partial t}\big)(t,R(t/2)\bar{\bx},x_3) +\frac{1}{2}\langle\Omega_0\big(\nabla_{\bar{\bx}}\psi)\big(t,R(t/2)\bar{\bx},x_3).$$
We used the fact that $\nabla_{\bar{\bx}}=R(t/2)\nabla_{\bar{\bQ}}$ as follows from the change of variables. From this, it also follows that 
$$\Delta_{\bar{\bQ}} = \langle\nabla_{\bar{\bQ}},\nabla_{\bar{\bQ}} =  \rangle=\langle\nabla_{\bar{\bx}},\nabla_{\bar{\bx}}\rangle = \Delta_{\bar{\bx}}.$$
These remarks establish the correspondence between solutions to \eqref{SE2} and \eqref{SE3}. To establish the correspondence between solutions to \eqref{SE1} and \eqref{SE2}, we implement \eqref{CT2} as a unitary transformation.  This case was dealt with in \cite{Gz2}, here we quote from that work. Beware of the changes in notation.

Again we use the proposal \eqref{QT1}, but with \eqref{CT2} in the exponent. So, suppose that $\varphi(t,\bxi)$ is a solution to \eqref{SE3} with initial data $\psi_0(\bxi),$ and put
\begin{equation}\label{QT2}
\phi(t,\bQ) = ( U_t\varphi)(t,\bQ) = \frac{1}{(2\pi)^{3/2}} \int e^{iF(\bQ,\bet,t)/\hbar}\phi(t,\bet)d\bet.
\end{equation}
Again, the transform can be explicitly computed. The result is:
\begin{equation}\label{QT2.1}
\phi(t,Q) = e^{i(f(\bQ,t) + A(t))/\hbar}\varphi(t, \bQ-\bQ_{nh}(t)).
\end{equation}
This transformation is unitary and besides the shift relative to $\bQ_{nh}(t),$ the new wave function acquires a global phase, which does not affect the normalization, but it does affect the computation of transition rates due to the perturbation as shown in \cite{Gz2}.
Where $A(t)$ was introduced in \eqref{act}, and we put $f(t,\bQ)=\langle\bQ-\bQ_{nh}(t)),\bP_{nh}(t)\rangle.$ It takes but a simple computation to verify that under $U_t$
$$\bxi\varphi(t,\bxi) \rightarrow (\bQ-\bQ_{nh}(t))\phi(t,\bQ)$$
$$-i\hbar\nabla_{\bxi}\varphi(t,\bxi) \rightarrow \big(-i\hbar\nabla_{\bQ}  + m\dot{\bQ}_{nh}(t)\big)\phi(t,\bQ).$$ 
This is the quantized version of \eqref{cteq2.1}. Observe that in the Hamiltonians \eqref{ch3} and \eqref{qh3}, the degrees of freedom are separated, and note as well that the generating function is additive (a sum of generating functions for each degree of freedom). Therefore to verify that $\phi(t,\bQ)$ satisfies \eqref{SE2} if $\varphi(t,\bxi)$ satisfies \eqref{SE3}, it suffices to do so for each degree of freedom. But this is explicitly carried out in \cite{Gz2} where it is carried out in detail.

To sum up, we have proved that, for a given initial state $\psi(0),$ we have: 
\begin{equation}
i\hbar\frac{\partial \varphi}{\partial t} = H_3\varphi \Longrightarrow i\hbar\frac{\partial \phi}{\partial t}H_2\phi \Longrightarrow i\hbar\frac{\partial \psi}{\partial t}H_2\psi, \;\;\;\varphi(0)=\phi(0)=\psi(0)
\end{equation}
whenever $\psi,$ $\phi$ and $\varphi$ are related by \eqref{QT2} and \eqref{QT1} respectively.
\section{The transformation of eigenstates}
Since the degrees of freedom are separated in $\bH_3,$ the eigenstates of definite energy are products of the eigenstates of each degree of freedom, and the total energy is the sum of the corresponding energies. In our case, we have:
\begin{gather}
 \varphi_{\bn,k}(\bar{\bxi},\xi_3) = \varphi_{n_1}(\xi_1)\varphi_{n_2}(\xi_2)e^{ikx_3/\hbar}\label{eigst3}\\
E_{\bn,k} = \hbar\omega(n_1+\frac{1}{2}) +  \hbar\omega(n_2+\frac{1}{2}) +\frac{\hbar^2k^2}{2m}. \label{eigvl3}
\end{gather}
We put $\bn=(n_1,n_2)$ as labels of the eigenstates (eigenvalues) of the 2-dimensional oscillator embedded in $\bH_3.$ Thus, the spectrum of $\bH_3$ has a discrete part embedded in a continuous part. Also, as usual, $\varphi_n$ is:
\begin{equation}\label{evec1}
\varphi_n(x) = \bigg(\frac{\alpha}{\pi^{1/2}2^nn!}\bigg)^{1/2}H_n(\alpha x)e^{-\frac{1}{2}x^2},
\end{equation}
Where $H_n(x)$ is the Hermite polynomial of degree $n,$ and $\alpha=\big(m^{3/2}\omega/\hbar\big)^{1/2}.$

The passage from \eqref{SE3} to \eqref{SE2} using \eqref{QT2.1}  involves each degree of freedom separately. And as the transformation only acts on the spatial part of the wave function, after applying \eqref{QT2.1}, the transform of \eqref{eigst3} is:

\begin{equation}\label{eigst2}
\begin{aligned}
\phi_{bn,k}(\bQ,t) &\\
=  e^{-itE_{\bn,k}/\hbar}&e^{i(f(\bQ,t) + A(t))/\hbar}\varphi_{n_1}(Q_1-Q_{1,nh}(t))\varphi_{n_2}(Q_2-Q_{2,nh}(t))e^{(-i\frac{k}{\hbar}(Q_3-Q_{3,nh}(t))}.
\end{aligned}
\end{equation}

The passage from the above solution to \eqref{SE2} to \eqref{SE1} is as in \eqref{QT2.1}, that is:
\begin{equation}\label{eigst3}
\begin{aligned}
&\psi_{\bn,k}(\bx,t) = \phi_{\bn,k}(R(t/2)\bx,t)\\
=  e^{-itE_{\bn,k}/\hbar}&e^{i(f(R(t/2)\bx,t) + A(t))/\hbar}\varphi_{n_1}(R_1(t)-R_{1,nh}(t))\varphi_{n_2}(R_2(t)-R_{2,nh}(t))e^{(-i\frac{k}{\hbar}(x_3-x_{3,nh}(t))}.
\end{aligned}
\end{equation}
To simplify the caligraphy, we introduced the notation:
\begin{equation}\label{sh1}
\begin{aligned}
R_i(t) &= \overline{(R(t/2)\bx)_i},\;\;\;\mbox{for} \;\;i=1,2.\\
x_{3,nh} &= Q_{3,nh}.
\end{aligned}
\end{equation}
The matrix $R(t)$ was introduced in \eqref{rot1}. To get rid of the shifted arguments in \eqref{sh1}, we make use of the summation formula:
\begin{equation}\label{summ}
H_n(u+v)=\sum_{k=0}^n{n\atopwithdelims()k}v^{n-k}H_k(u)..
\end{equation}
Invoke \eqref{evec1} to obtain:
\begin{equation}\label{evec2}
\varphi(u+v) = \sum_{k=0}^n A_{n,k}(\alpha v)^{n-k}\varphi_n(x), \;\;\;\mbox{with}\;\;\; A_{n,k}=\bigg(\frac{2^kk!}{2^nn!}\bigg)^{1/2}{n\atopwithdelims()k}.
\end{equation}
With this, we obtain $\phi_{\bn,k}(\bar{\bx})$ as a linear combination of products of the type 
$$\varphi_{k_1}\big((R(t/2)\bx)_1\big)\varphi_{k_2}\big((R(t/2)\bx)_2\big),\;\;\;0\leq k_1\leq n_1,\;0\leq k_2\leq n_2,$$
The last step in the chain consists of writing each of these products as linear combinations of products like $\varphi_{m_1}(\bx_1)\varphi_{m_1}(\bx_2)$ where $m_1+m_2=k_1+k_2.$ This will render $\psi(\bn,k)$ as a global phase multiplying a wave function which is a linear combination of eigenstates of energies less or equal that $E_{\bn,k}.$ The computation is carried out in considerable detail in \cite{Gz1}. We just quote the result.

\begin{equation}\label{entang}
\begin{aligned}
\varphi_{k_1}&\big((R(t/2)\bx)_1\big)\varphi_{k_2}\big((R(t/2)\bx)_2\big)\\
& =\sum_{l_1=0}^{k_1}\sum_{l_2=0}^{k_2}D(k_1,k_2,l_1,l_2,t){k_1\atopwithdelims()l_1}{n_2\atopwithdelims()l_2}s^{k_1}_1s^{k_2}_2\psi_{l_1+l_2}(x_1)\psi_{(k_1+k_2)-(l_1+l_2)}(x_2).
\end{aligned}
\end{equation}

\section{A particle under the action of oscillating magnetic fields and time-varying electric fields}
Here we consider two variations on the theme of \cite{dJ}, with notations somewhat different from theirs to make it consistent with our notations. 

\subsection{Case I: Magnetic field has fixed direction but is time dependent}.
This case is very similar to the case considered in Section 1, except that now $\bB=(0,0,B_3(t))^\dag.$  The classical and quantum Hamiltonians differ only in the fact that $B_3(t)$ is time-dependent, therefore, the cross-product matrix $\bOm(t)$ is time-dependent, and the rotations that it generates are a bit more elaborate.  The rotation matrix $R(t)$ introduced in \eqref{rot1}, is now defined by
\begin{equation}\label{rot3.0}
\frac{d}{dt}R(t) = \bOm(t)R(t).
\end{equation}
It is easy to verify that in this case
\begin{equation}\label{rot3.1}
R(t) = \begin{pmatrix}
		cos(A(t)) & -sin(A(t)) & 0\\
		sin(A(t)  & cos(A(t)) & 0\\
		0  & 0  & 1\end{pmatrix} \;\;\;\mbox{where}\;\;\;\;A(t)=\int_0^t\omega(s)ds.
\end{equation}
This is an easy consequence of the fact that the axis of rotation is kept fixed and that the structure of  $\bOm(t)$ is such that it commutes with itself at different times. This time, since $R(t)$ commutes with $\bOm(t),$ makes it easy to translate the arguments developed above to this case as well. The difference is that now the cyclotron frequency in \eqref{ch3} and \eqref{qh3} is time-dependent and we are left with a dynamics described by an equation of the Hill type. See \cite{MW} for example.

\subsection{Case II: Magnetic field rotates about a fixed axis in space}

The classical Hamiltonian of the system is

\begin{equation}\label{ch4}
H_4(\bx,\bp) = \frac{1}{2m}\bigg(\bp-\bA(\bx,t)\bigg)^2 -q\langle\bx,\bE_0(t)\rangle.
\end{equation}
In \cite{dJ} the last term is absent, and an electric field eventually comes up rearranging their Hamiltonian. Such last term is the analog of the electric field that appears in the passage from \eqref{ch1} to \eqref{ch2}. But we might as well have the $\bE_0(t)$ because we already know how to solve the resulting problem as shown in \cite{Gz2} and extended above. Note that in this case, we might as well consider that the electric field $-\partial \bA/\partial t,$ which is linear in $\bx,$ to be subsumed as part of $\bE_0(t).$ Here we prove that the Hamiltonian \eqref{ch4} is also equivalent to \eqref{ch2}.

To further specify this system, we again set $\bA=\frac{1}{2}\bB(t)\times\bx,$ where $B(t)=R(t)\bB(0),$ where $R(t)$ describes a rotation about the $\hat{\bk}$-axis whose phase has been adjusted so that $\bB(0)=(B_1,0,B_3).$  Explicitly:

\begin{equation}\label{rot3}
\bB(t) = \begin{pmatrix}
		cos(\alpha t) & -sin(\alpha t) & 0\\
		sin(\alpha t)  & cos(\alpha t) & 0\\
		0  & 0  & 1\end{pmatrix} \begin{pmatrix} B_1\\ 0\\B_3\end{pmatrix}.
\end{equation}
The infinitesimal generator of this rotation group is the matrix $\bLambda$ given by

\begin{equation}\label{infgen2}
\bLambda = \begin{pmatrix}
		0 & -\alpha &  0 \\
		\alpha & 0 &   0\\
		0 & 0  & 0\end{pmatrix} .
\end{equation}
The analog of the matrix $\bOm$ introduced in \eqref{ang1} is now

\begin{equation}\label{ang4}
\bOm_1(t) = \begin{pmatrix}
		0 & -\omega_3 &  B_1\sin(\alpha t) \\
		B_3 & 0 &   -B_1\cos(\alpha t) \\
		-B_1\sin(\alpha t)  & B_1\cos(\alpha t)  & 0\end{pmatrix} ,\;\;\;\;\;
\bOm_1(0) = \begin{pmatrix}		
                 0 &  -B_3  &  0 \\
		B_3 & 0 &   -B_1\ \\
		0 & B_1  & 0\end{pmatrix} 
\end{equation}
We have introduced the cyclotronic frequencies 
\begin{equation}\label{cyc}
\omega_i = \frac{qB_i}{mc}.
\end{equation}

We leave it to the reader to verify that $R(t)\bOm(t)R(-t) = \bOm(0).$ This means that $\bOm(t)$ satisfies the Euler equation $\dot{\bOm} + [\bLambda,\bOm]=0$ with initial condition $\bOm(0).$
We pass to a coordinate system in which the horizontal component of the magnetic field is constant using a time-dependent canonical transformation generated by

\begin{equation}\label{CT3}
F_(\bx,\bP) = \langle R(-t)\bx,\bP\rangle.
\end{equation}

 The new canonical variables are:
 \begin{equation}\label{ct3.1}
 \bQ = \nabla_{\bP}F = R(-t)\bx,\;\;\;\mbox{and}\;\;\;\bp = \nabla_{\bx}F\Rightarrow \bP=R(-t)\bp.
 \end{equation}
 Similarly, invoking the invariance of the scalar products under rotations, and after some simple arithmetics, the new Hamiltonian function is
 $$H_5(\bQ,\bP) = H_4(R(t)\bQ,R(t)\bP) + \bigg(\frac{\partial F}{\partial t}\bigg)(R(t)\bQ,\bP).$$
 Doing the substitutions we obtain:
 \begin{equation}\label{ch5}
 H_5(\bQ,\bP) = \frac{1}{2m}\bP^2 -  \langle\bP,\big(\frac{1}{2}\bOm_1(0) + \bLambda\big)\bQ\rangle + \frac{m}{2}\langle\bQ,\bOm_1^t(0)\bOm_1(0)\bQ\rangle - q\langle\bQ,\bE_1(t)\rangle.
 \end{equation}
 We put $bE_1(t)=R(t)\bE_0(t).$ Let us write $\bM=\frac{1}{2}\bOm_1(0) + \bLambda$ Note that $\bM^t=-\bM,$ therefore $G(t)=\exp(\bM t)$ is a rotation group about the axis
 $$\bn=\left(B_1/2,0,(\alpha+b_3/2)\right)^\dag/\big((B_1^2/2)+(\alpha+B_3/2)^2\big)^{1/2}$$
 with speed $\theta=\big((B_1/2)^2+(\alpha+B_3/2)^2\big)^{1/2}.$ Now we repeat the procedure to eliminate the second term in the right-hand side of \eqref{ch5}. Considering the
 transformation generated by:
 
 \begin{equation}\label{CT4}
 F_(\bQ,\bp') = \langle e^{t\bM}\bQ,\bp'\rangle.
\end{equation}

we obtain, that in the coordinates $(\bx',\bp'),$ the new Hamiltonian looks like
  \begin{equation}\label{ch5}
 H_5(\bx',\bp') =  \frac{1}{2m}(\bp')^2  + \frac{m}{2}\langle\bQ,\bOm_1^t(0)\bOm_1(0)\bQ\rangle - q\langle\bQ,\bE_1(t)\rangle.
 \end{equation}
 
 As in the previous case, this Hamiltonian describes an oscillator with a time-dependent frequency. We end this section with the following remark. One might be tempted to carry out a time-dependent rotation that aligns the magnetic field to the $\hat{\bk}-$axis. And then do as we did in Section 2 to compensate the magnetic field away.  But as shown in the first case, this will lead to an oscillator with a time-dependent frequency. 
 
 \section{Final remarks}
 To sum up, only when the magnetic field is static, it can be compensated away by a rotating coordinate system, in which the motion, or the time evolution, is canonically equivalent to that of a harmonic oscillator. When there is also a static electric field, the resulting motion is equivalent to that of a harmonic oscillator subject to a spatially homogeneous time-dependent force. In this case, the solution of the Schr\"{o}dinger equation acquires a global phase, and as mentioned, the eigenstates of the quantum system happen to be entangled states of simple harmonic oscillators.
 
 When the magnetic field is time-dependent, the system is canonically (or unitarily) equivalent to an oscillator with time-varying frequency. As there does not exist a general solution for this case, one must resort to different types of approximations. See \cite {K} for early work in this direction. 
 
 {\bf Declaration of competing interests} I have no competing interests to declare, no funding to report and this work complies the highest standars of ethical conduct.

\end{document}